\documentclass[twocolumn,aps,amsmath,amssymb,floatfix,superscriptaddress]{revtex4}
\pdfoutput=1 %This is for the ArXiv submission only
\usepackage{graphicx,eucal}
\usepackage[all]{xy}

\usepackage[colorlinks=true, urlcolor=blue, anchorcolor=blue, citecolor=blue,filecolor=blue,linkcolor=blue,menucolor=blue]{hyperref}

\newcommand{\ii}{{\rm  i}}

\begin{document}
\title{Infection process near criticality: Influence of the initial condition}

\author{P.~L.~Krapivsky}
\affiliation{Department of Physics, Boston University, Boston, Massachusetts 02215, USA}

\begin{abstract}
We investigate how the initial number of infected individuals affects the behavior of the critical susceptible-infected-recovered process. We analyze the outbreak size distribution, duration of the outbreaks, and the role of fluctuations. 
\end{abstract}

\maketitle

\section{Introduction}
\label{sec:intro}

Epidemics greatly affect Homo sapiens \cite{McN,O,Benedict} and other animals. Epidemic modeling goes back to Bernoulli who modeled the spread of smallpox \cite{Ber,Ber-history}. The methods and concepts developed in this field have been applied to modeling recent human epidemics like HIV, H1N1 Swine Flu, Zika Virus, and COVID-19, and also to modeling non-biological processes like the spread of rumors or cultural fads, diffusion of innovations, computer viruses, etc. (see \cite{MT,Lud,CFL,Volovik} and references therein).

Epidemics often operate close to the critical regime. Subcritical epidemics are the most numerous, but they quickly die out. Supercritical epidemics can kill a finite fraction of the population thereby destroying the environment that the virus needs to thrive and multiply. Epidemics close to the critical regime are optimal in a view of never-ending competition between the hosts and the viruses. Non-biological factors play an increasingly important role in the spread of epidemics in human society. On one side, modern human society is much more interconnected than ever --- this facilitates the spread of infections and leaves no hope for their total eradication. On the other side, advances in technology and wealth help to devise and employ the containment measures that suppress supercritical epidemics.  

Mathematical models of epidemics are over-simplified. In physics, for instance, we know that electrons are identical. In contrast, organisms are different, even twins are different. This heterogeneity is rarely taken into account, and it is far from clear how to model it in a reasonable way. In physics, we usually rely on binary and symmetric interactions. Both these features are questionable in the realm of epidemics. Other realistic features are also mostly ignored. However, the populations where the epidemics spread are usually very large and the lore from statistical physics tells us that in large systems qualitative behaviors can be predicted even if one greatly simplifies the model. This is especially true in critical regimes. 

The very concept of the critical regime comes from epidemic modeling. This concept clearly emerges from the well-known susceptible-infected-recovered (SIR) process \cite{McK,KMcK,May,Siam,Murray}, a toy model that mimics the spread of infection. According to the rules of the SIR process, infected individuals recover (become immune or die) with equal rates  and every infected individual transmits a disease to every susceptible individual with the rate $R_0/N$, where $N$ is the population size.  Thus on average, each infected individual spreads the infection to $R_0$ individuals before recovery. Therefore the behavior of the SIR process greatly depends on whether the reproduction number $R_0$ is smaller or larger than the recovery rate which we set to unity. When $R_0<1$,  i.e., for subcritical SIR processes,  outbreaks quickly end, namely just a few individuals catch the disease. For supercritical SIR processes ($R_0>1$), the outbreak may affect only a few individuals, e.g. starting from a single infected individual the size of the outbreak is finite with probability $R_0^{-1}$. With complementary probability, $rN+O(\sqrt{N})$ individuals catch the disease before the outbreaks dies out;  the fraction $r=r(R_0)$ is implicitly determined by 
\begin{equation}
\label{rR}
r+e^{-R_0 r} = 1 
\end{equation}

Huge outbreaks killing finite fractions of the population continue to devastate animal species. They also used to decimate human societies \cite{McN,O,Benedict}, e.g., the Black Death killed about 50\% of the European population \cite{Benedict}. Preventive and containment measures such as quarantine, improved hygiene, etc. suppress supercritical infectious diseases and often drive them to a critical situation. This critical state is effectively self-organized. Indeed, suppressing the disease to the subcritical regime may be possible but costly and psychologically difficult to maintain when the number of newly infected starts to decrease exponentially. Therefore, if the outbreak is not quickly and completely eradicated, the containment measures are relaxed and the system may return to the supercritical stage, the disease gets again out of control, so the containment  measures are tightened driving the system back to the subcritical state. It would be interesting to devise a self-organized process of the spread of infection with dynamics similar to the critical SIR process. In this paper, however,  we merely consider the critical SIR process with many initially infected individuals. 

The SIR processes are often treated using a deterministic framework \cite{McK,KMcK,May,Siam,Murray}. This framework can be applied to the supercritical regime where it gives e.g. the simplest derivation of  Eq.~\eqref{rR}.  Stochastic effects are unavoidable, however, for the critical and subcritical SIR processes (see \cite{Bailey50,Bailey,AB,book}). When the population of susceptible is finite, finite-size corrections become important, particularly for the critical SIR process \cite{rr,ML,bk,KS,Gordillo,Hofstad,bk12}. 

In this paper, we study the critical SIR process and hence we employ stochastic methods. We consider finite populations. The population size is assumed to be large, $N\gg 1$.  We focus on the situation when the initial number of infected individuals is also large: $k\gg 1$. This may seem unrealistic as most epidemics start with a single infected individual. Our goal, however, is to understand the behavior of the SIR process that has begun at the supercritical regime, exhibited an exponential growth regime, and was subsequently suppressed (by preventive and containment measures) to the critical SIR. Ignoring the earlier regime yields the critical SIR process with a certain large number $k$ of initially infected individuals. 

The same critical SIR process with a large number of initially infected individuals has been recently studied, using large-scale simulations and scaling arguments, by Radicchi and Bianconi \cite{Ginestra}. We rely on exact calculations and asymptotic analyses. Our analytical and asymptotic predictions qualitatively agree with simulation results \cite{Ginestra}. Some quantitive discrepancies suggest trying slightly different scaling fits that are a little simpler than the fits used in \cite{Ginestra}. The chief reason for subtle behaviors are algebraic tails; the average size and duration of outbreaks are especially sensitive to these tails. 

The chief outcome of the epidemic is the outbreak size $n$. The full description of this random quantity is provided by the probability $A_n(k, N)$ that starting with $k$ infected individuals in the population of size $N$,  exactly $n$ individuals catch the infection before the epidemic stops. We examine this probability distribution. In particular, we study the behavior of the average and variance
\begin{equation}
\label{def:E-Var}
\mathbb{E}_k(N)=\langle n\rangle, \qquad \mathbb{V}_k(N)=\langle n^2\rangle- \langle n\rangle^2
\end{equation}
We argue that in the $N\to\infty$ limit, these quantities depend on a single variable $\kappa=k/N^{1/3}$. More precisely, they exhibit the following scaling behaviors
\begin{equation}
\label{av-var-scaling} 
\mathbb{E}_k(N)  = N^{2/3}\mathcal{E}(\kappa), \quad \mathbb{V}_k(N)=N^{4/3}\mathcal{V}(\kappa)
\end{equation}
We shall show that the scaled distributions have simple extremal behaviors 
\begin{subequations}
\begin{equation}
\label{av-extreme}
\mathcal{E}(\kappa) =
\begin{cases}
C_1\kappa        &\text{when}\quad \kappa\ll 1\\
\sqrt{2\kappa}   &\text{when}\quad \kappa\gg 1
\end{cases}
\end{equation}
and 
\begin{equation}
\label{var-extreme}
\mathcal{V}(\kappa) =
\begin{cases}
C_2\kappa        &\text{when}\quad \kappa\ll 1\\
\sqrt{2/\kappa}   &\text{when}\quad \kappa\gg 1
\end{cases}
\end{equation}
\end{subequations}

Considerable insight into the behavior of finite systems can be gained from the analysis of the infinite-population limit. In this limit, the critical SIR process is equivalent to the critical branching process. This classical process is traditionally studied starting with a single infected individual. We allow an arbitrary number $k$ of initially infected individuals and derive several exact results in the infinite-population limit.  

The outline of this paper is as follows. In Sec.~\ref{sec:CBP}, we consider the infinite-population limit, present exact results for the outbreak size distribution, and show that exact results approach a simple scaling form in the most interesting situation when $k\gg 1$. We then analyze the critical SIR process in a population with $N\gg 1$ individuals. We study the distribution of the size of an outbreak using scaling and heuristic arguments (Sec.~\ref{sec:SIR}) and asymptotically exact analysis (Sec.~\ref{sec:EA}). In particular, in these sections we derive \eqref{av-extreme}--\eqref{var-extreme}. In Sec.~\ref{sec:SIR-time}, we investigate the duration of outbreaks. Several technical calculations are relegated to the Appendices~\ref{ap:record}--\ref{ap:time}.

\section{infinite-population limit: Outbreak Size Distribution}
\label{sec:CBP}

In the infinite-population limit, the SIR process reduces to the branching process \cite{feller,teh,athreya04,branch,vatutin}. Branching processes involve duplication and death. Symbolically, 
\begin{displaymath}
\xymatrix{P+P                & \\
               P \ar[u] \ar[r]   & \emptyset}
\end{displaymath}
For the critical branching process, the rates of duplication and death are equal. 

Branching processes have numerous applications, e.g., they mimic cell division and death; for adult organisms, the critical branching process is appropriate as the number of cells remains (approximately) constant. 

We begin with a classical setting when one individual was initially infected. Let $A_n$ be the probability that exactly $n$ individuals catch the infection before the epidemic is over.  With probability $\tfrac{1}{2}$, the initially infected individual joins the population of recovered before infecting anyone else, so $A_1=\tfrac{1}{2}$. Further, $A_2=\tfrac{1}{2}A_1^2$ since at the first step a new individual must get infected, and then both must recover without spreading infection. Proceeding along these lines we arrive at the recurrence 
\begin{equation}
\label{An_rec}
A_n = \frac{1}{2}\sum_{i+j=n}A_iA_j+\frac{1}{2}\,\delta_{n,1}
\end{equation}
reflecting that the first infection event creates two independent infection processes \cite{teh}.  A solution to \eqref{An_rec} is found by introducing the generating function
\begin{equation}
\label{An_gen}
A(z) = \sum_{n\geq 1}A_n z^n
\end{equation}
converting the recurrence \eqref{An_rec} into a quadratic equation 
\begin{equation}
\label{2A:eq}
2 A(z) = [A(z)]^2 + z
\end{equation}
whose solution reads
\begin{equation}
\label{Az}
A(z)  = 1 - \sqrt{1-z}
\end{equation}
Expanding $A(z)$ in powers of $z$ we find
\begin{equation}
\label{An_sol}
A_n = \frac{1}{\sqrt{4\pi}}\, \frac{\Gamma\left(n-\tfrac{1}{2}\right)}{\Gamma(n+1)}
\simeq \frac{1}{\sqrt{4\pi}}\,n^{-3/2}
\end{equation}
In particular, the probabilities $A_n$ are given by 
\begin{equation*}
 \tfrac{1}{2},\tfrac{1}{8},\tfrac{1}{16},\tfrac{5}{128},\tfrac{7}{256},\tfrac{21}{1024},\tfrac{33}{2048},\tfrac{429}{32768},\tfrac{715}{65536},\tfrac{2431}{262144},\tfrac{4199}{524288}
\end{equation*}
for $n=1,\dots,11$. 

Generally when the critical branching process begins with $k$ initially infected individuals, infection processes originated with each individual are independent. Hence the probability $A_n^{(k)}$ that exactly $n$ individuals catch the infection before the epidemic is over can be expressed via the probabilities $A_m \equiv A_m^{(1)}$ corresponding to the classical situation with one initially infected individual: 
\begin{equation}
\label{An-k}
A_n^{(k)} = \sum_{i_1+\ldots+i_k=n}A_{i_1}\ldots A_{i_k}
\end{equation}
The generating function
\begin{equation}
\label{gen-k-def}
A^{(k)}(z) = \sum_{n\geq k}A_n^{(k)} z^n
\end{equation}
is therefore
\begin{equation}
\label{gen-k}
A^{(k)}(z) = [A(z)]^k = \left\{1 - \sqrt{1-z}\right\}^k
\end{equation}
This generating function encapsulates all $A_n^{(k)}$.

Let us first look at the probabilities $A_n^{(k)}$ for small $k$. Needless to say, $A_n^{(k)}=0$ when $n<k$. When $n\geq k$, one can express $A_n^{(k)}$ through the probabilities \eqref{An_sol}. Here are a first few explicit formulas
\begin{equation*}
\begin{split}
&A_n^{(2)}   = 2A_n\\
&A_n^{(3)}  = 4A_n - A_{n-1}\\
&A_n^{(4)}  =8A_n - 4A_{n-1}\\
&A_n^{(5)}  =16A_n - 12A_{n-1} + A_{n-2}\\
&A_n^{(6)}  =32A_n - 32A_{n-1} + 6A_{n-2}\\
&A_n^{(7)}  =64A_n - 80A_{n-1} + 24A_{n-2}-A_{n-3}\\
&A_n^{(8)}  =128A_n - 192A_{n-1} + 80A_{n-2}-8A_{n-3}\\
&A_n^{(9)}  =256A_n - 448A_{n-1} + 240A_{n-2}-40A_{n-3}+ A_{n-4}
\end{split}
\end{equation*}
suggesting the general representation of $A_n^{(k)}$ as a sum
\begin{equation}
\label{Ank-sum}
A_n^{(k)}=\sum_{p=0}^{\lfloor \frac{k-1}{2}\rfloor} (-1)^p \binom{k-1-p}{p} 2^{k-1-2p}A_{n-p}
\end{equation}
Here $\lfloor x\rfloor$ is the largest integer $\leq x$ and $ \binom{a}{b}$ are binomial coefficients. Massaging the sum in Eq.~\eqref{Ank-sum} one obtains a neat final formula 
\begin{equation}
\label{Ank-gamma}
A_n^{(k)}=\frac{k}{2^{2n-k}}\,\frac{\Gamma(2n-k)}{ \Gamma(n+1-k)\Gamma(n+1)}
\end{equation}
The derivation of Eq.~\eqref{Ank-gamma} from \eqref{Ank-sum} is outlined in Appendix~\ref{ap:record}, where we also present a more straightforward derivation of Eq.~\eqref{Ank-gamma}  from the generating function \eqref{gen-k}. 

\begin{figure}
\centering
\includegraphics[width=7.89cm]{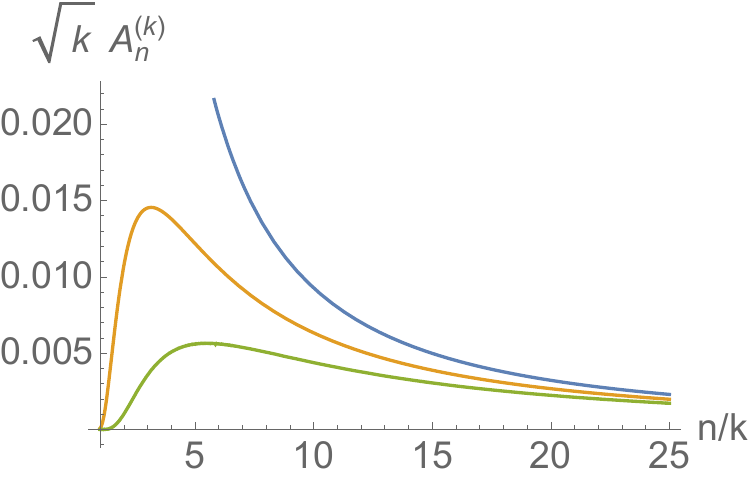}
\caption{Top to bottom:  $\sqrt{k}\,A_n^{(k)}$  versus $n/k$ for $k=1, 16, 30$. The asymptotic behavior is $(4\pi)^{-1/2} (n/k)^{-3/2}$  in agreement with Eq.~\eqref{An-k-asymp}. }
\label{Fig:An-k}
\end{figure}

Note that substituting \eqref{An_sol} into the recurrence \eqref{An-k} one can directly compute
\begin{equation}
\label{Akk}
\begin{split}
&k\geq 1: \quad A_k^{(k)} =\frac{1}{2^k}  \\
&k\geq 2: \quad A_{k+1}^{(k)}=\frac{k}{2^{k+2}} \\
&k\geq 3: \quad A_{k+2}^{(k)}=\frac{k}{2^{k+3}}+ \frac{k(k-1)}{2^{k+5}}
\end{split}
\end{equation}
These results are recovered from the general solution \eqref{Ank-gamma} thereby providing a consistency check. As another consistency check we note that \eqref{Ank-gamma} agrees with normalization 
\begin{equation}
\label{norm}
\sum_{n\geq k} A_n^{(k)}=1
\end{equation}

The sequence $A_n^{(k)}$ has a single peak located at $n=k$ when $k\leq 4$, while for $k\geq 5$ the peak is at $n=\nu(k)>k$, see Fig.~\ref{Fig:An-k}. The sequence $A_n^{(k)}$ grows from $A_k^{(k)}=2^{-k}$ to the maximum
\begin{equation}
\label{An-k-max}
b(k):=\text{max}\{A_n^{(k)}|\, n\geq k\}
\end{equation}
at $n=\nu(k)$, then decays and eventually approaches 
\begin{equation}
\label{An-k-asymp}
A_n^{(k)}\simeq \frac{k}{\sqrt{4\pi}}\,n^{-3/2} 
\end{equation}
This asymptotic behavior is straightforwardly deduced from  \eqref{Ank-gamma}. Using the general solution \eqref{Ank-gamma} together with the Stirling formula one deduces the behaviors of $b(k)$ and $\nu(k)$ in the $k\to\infty$ limit:
\begin{equation}
\label{nu-a}
\nu(k)\simeq \tfrac{1}{6}k^2, \qquad b(k)\simeq B k^{-2}
\end{equation}
with
\begin{equation}
\label{C:def}
B=3e^{-3/2}\sqrt{\frac{6}{\pi}}= 0.92508197882\ldots
\end{equation}

The general solution \eqref{Ank-gamma} approaches the scaling form 
\begin{equation}
\label{Ank-scaling}
A_n^{(k)}=\frac{4}{k^2}\,\Phi(\mu), \quad \Phi(\mu)=\pi^{-1/2} \mu^{-3/2}\,e^{-1/\mu}
\end{equation}
in the scaling limit
\begin{equation}
\label{mu-scaling}
n\to\infty, \quad k\to \infty, \quad \mu=\frac{4n}{k^2}= \text{finite}
\end{equation}

\begin{figure}
\centering
\includegraphics[width=7.89cm]{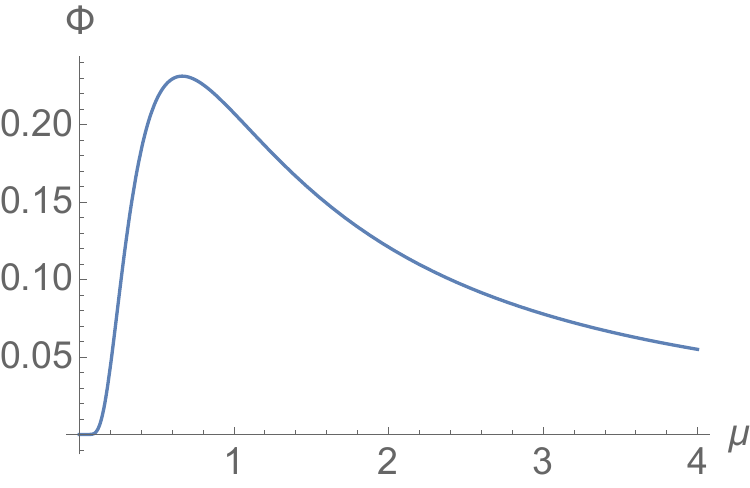}
\caption{The scaled distribution $\Phi(\mu)$ versus the scaled outbreak size $\mu$ given by Eq.~\eqref{Ank-scaling}. }
\label{Fig:Phi}
\end{figure}

The scaled distribution $\Phi(\mu)$ has a single peak (Fig.~\ref{Fig:Phi}) and it vanishes faster than any power of $\mu$ when $\mu\to 0$. We also note that from Eqs.~\eqref{Ank-scaling}--\eqref{mu-scaling} one easily recovers \eqref{nu-a}--\eqref{C:def}. Due to the algebraic tail \eqref{An-k-asymp}, the moments defined by $\langle n^a\rangle = \sum_{n\geq k} n^a A_n^{(k)}$ diverge when $a\geq \frac{1}{2}$. To perform the summation  in the $a<\frac{1}{2}$ range where the moments converge, let us first consider modified moments with $n^a$ replaced by $\Gamma(n+1)/\Gamma(n+1-a)$. Such moments admit an analytical expression 
\begin{equation}
\label{moments-a-k}
\sum_{n\geq k} \frac{\Gamma(n+1)}{\Gamma(n+1-a)}\, A_n^{(k)}=\frac{\Gamma(1-2a)}{\Gamma(1-a)}\, \frac{\Gamma(1+k)}{\Gamma(1-2a+k)}
\end{equation}
This result has been established  by substituting $A_n^{(k)}$ given by \eqref{Ank-gamma} into Eq.~\eqref{moments-a-k} and computing the 
sum using identities that can be found, e.g., in Ref.~\cite{Knuth}. 

When $k\gg 1$, the modified moments are asymptotically equal to the regular moments since $\frac{\Gamma(n+1)}{\Gamma(n+1-a)}\to n^a$ for $n\geq k\gg 1$. Therefore Eq.~\eqref{moments-a-k} simplifies to
\begin{equation}
\label{mom-a-k}
\langle n^a\rangle=\sum_{n\geq k} n^a A_n^{(k)}\simeq \frac{\Gamma(1-2a)}{\Gamma(1-a)}\,  k^{2a}
\end{equation}
when $a<\frac{1}{2}$. In the complementary $a\geq \frac{1}{2}$, the moments $\langle n^a\rangle$ remain finite for finite populations, but diverge as $N\to\infty$. This divergent leading behavior of the moments in finite populations is computed in the next section; in the simplest case of $k=1$, it is given by Eq.~\eqref{mom-a-k-N}.

\section{Outbreak Size Distribution: Scaling Analysis}
\label{sec:SIR}

Consider the critical SIR process with $k$ initially infected individuals, but in a finite population. Denote by $N$ the size of the population and by $A_n(k; N)$ the probability that the size of the outbreak is $n$. In the infinite-population limit,  $A_n(k; \infty)\equiv A_n^{(k)}$ is given by \eqref{Ank-gamma}; it approaches the scaling form \eqref{Ank-scaling}--\eqref{mu-scaling} when $k\gg 1$. 

The probability distribution $A_n(N)\equiv A_n(1; N)$ for the critical SIR process starting with a single infected individual has been investigated in Refs.~\cite{ML,bk,KS,Gordillo,Hofstad,bk12}. The probability distribution $A_n(N)$ acquires a scaling form 
\begin{equation}
\label{AAF}
A_n(N)=  A_n\, \mathcal{A}(\nu)
\end{equation}
in the scaling limit 
\begin{equation}
\label{scaling} 
n\to\infty, ~~N\to\infty, ~~\nu=\frac{n}{N^{2/3}}=\text{finite}
\end{equation}

This scaling was proposed in \cite{bk}. Simulations of very large systems with $N$ comparable to the current human population \cite{bk,bk12} are well fitted by the scaling form \eqref{AAF}--\eqref{scaling}. Kessler and Shnerb \cite{KS} derived the scaling function $\mathcal{A}(\nu)$; similar scaling functions have been rigorously studied in Refs.~\cite{Gordillo,Hofstad}. 

Recall that the moments $\langle n^a\rangle$ with $a\geq \frac{1}{2}$ diverge in the infinite-population limit. For finite populations, these moments are finite and the scaling \eqref{AAF}--\eqref{scaling} implies the following leading asymptotic behavior
\begin{equation}
\label{mom-a-k-N}
\langle n^a\rangle\simeq
\begin{cases} 
(9\pi)^{-1/2}\ln N             & a = \frac{1}{2}\\
C_a N^\frac{2a-1}{3}     & a > \frac{1}{2}
\end{cases}
\end{equation}
with
\begin{equation}
\label{Ca}
C_a = \int_0^\infty \frac{d\nu}{\sqrt{4 \pi \nu^3}}\,\,\nu^a\,\mathcal{A}(\nu)
\end{equation}

Two important properties of the epidemics are the average and variance of the size of outbreaks. These quantities are defined by Eq.~\eqref{def:E-Var}. 
When $k=1$, the average and the variance scale with population size as 
\begin{equation}
\label{EV:CC}
\mathbb{E}_1(N) \simeq C_1 N^{1/3}, \qquad \mathbb{V}_1(N) \simeq C_2 N
\end{equation}
Indeed, Eq.~\eqref{mom-a-k-N} shows that $\langle n^2\rangle \gg \langle n\rangle^2$ when $N\gg 1$, so Eq.~\eqref{EV:CC} follows from \eqref{mom-a-k-N}, and the amplitudes are given by \eqref{Ca}.  We have computed the amplitudes using the exact expression  \cite{KS} for the scaling function to give  
\begin{equation}
\label{CC:12}
C_1=1.4528\ldots, \quad  C_2\approx 3.99
\end{equation}

In the general case, the distribution $A_n(k,N)$ depends on three variables. The interesting range is
\begin{equation}
\label{nNkN}
n\sim N^{2/3}, \quad k\sim N^{1/3}
\end{equation}
The first scaling law follows from Eq.~\eqref{scaling}, the second is an outcome of Eqs.~\eqref{mu-scaling} and \eqref{scaling}. In the scaling region \eqref{nNkN},  the distribution $A_n(k,N)$ is expected to acquire a scaling form
\begin{equation}
\label{ANG}
A_n(k,N) =  N^{-2/3}\, \mathcal{A}(\kappa,\nu)
\end{equation}
with $\nu=n/N^{2/3}$, see Eq.~\eqref{scaling}, and the scaled initial number of infected individuals $\kappa=k/N^{1/3}$. More precisely, \eqref{ANG} should hold in the scaling limit \eqref{scaling} and 
\begin{equation}
\label{kN-scaling} 
k\to\infty, ~~N\to\infty, ~~\kappa=\frac{k}{N^{1/3}}=\text{finite}
\end{equation}
The normalization condition, $\sum_{n\geq k}A_n(k,N)=1$, gives $\int_0^\infty d\nu\, \mathcal{A}(\kappa,\nu)=1$ and explains the pre-factor in \eqref{ANG}. 

The two-variable distribution $\mathcal{A}(\kappa,\nu)$ is unknown, so let us discuss the average outbreak size which is the basic quantity with simpler scaling behavior. The average outbreak size $\mathbb{E}_k(N)$ depends on two variables and its conjectural scaling behavior is
\begin{equation}
\label{av-scaling} 
\mathbb{E}_k(N)  = N^{2/3} \mathcal{E}(\kappa)
\end{equation}
The two scaling behaviors, \eqref{ANG} and \eqref{av-scaling}, are compatible if $\mathcal{E}(\kappa) = \int_0^\infty d\nu\, \nu\, \mathcal{A}(\kappa,\nu)$. 

Let us probe the extremal behaviors of the scaled distribution $\mathcal{E}(\kappa)$. To establish the small $\kappa$ asymptotic, we note that in the $k\ll N^{1/3}$ regime, the infectious processes generated by each initially infected individual are mutually independent. Thus
\begin{equation}
\label{NkN-asymp} 
\mathbb{E}_k(N) \simeq C_1\, k\, N^{1/3} \qquad\text{when}\quad  k\ll N^{1/3}
\end{equation}
Comparing \eqref{av-scaling} and \eqref{NkN-asymp} we obtain $\mathcal{E}(\kappa)\simeq C_1\kappa$ as $\kappa\to 0$ as we stated in Eq.~\eqref{av-extreme}. We  also fix the amplitude $C_1$ appearing in Eq.~\eqref{av-extreme}, it is given by \eqref{CC:12}. 

To appreciate the large $\kappa$ behavior of $\mathcal{E}(\kappa)$ we mention that $\mathbb{E}_N(N) = N$. This suggests that $\mathcal{E}(N^{2/3}) \sim N^{1/3}$, from which $\mathcal{E}(\kappa)\sim \sqrt{\kappa}$ as $\kappa\to \infty$. This argument is heuristic. In Sec.~\ref{sec:EA}, we derive the large $\kappa$ asymptotic stated in Eq.~\eqref{av-extreme}.  In Appendix \ref{ap:det}, we present another elementary derivation based on the observation that the behavior in the $\kappa\to \infty$ limit is essentially deterministic. The deterministic analysis is significantly simpler than the exact approach, but it is not suitable for studying fluctuations. 

Simulation results \cite{Ginestra} are in a fairly good agreement with the linear behavior of the scaled average outbreak size in the small $\kappa$ limit: $\mathcal{E}(\kappa) = C_1\kappa$ with $C_1\approx 1.5$ in simulations,  the analytical prediction is $C_1=1.4528\ldots$. In the large $\kappa$ limit, numerical data in \cite{Ginestra} were fitted to $\sqrt{\kappa}$ up to a logarithmic correction. 

We now turn to the variance of the size of the outbreaks. For sufficiently small $k$, we rely again on the mutual independence of $k$ infectious processes to deduce
\begin{equation}
\label{Var-asymp} 
\mathbb{V}_k(N)  \simeq C_2\, k\, N \qquad\text{when}\quad  k\ll N^{1/3}
\end{equation}
The scaling region is given by \eqref{kN-scaling}. The natural scaling behavior of the variance compatible with \eqref{Var-asymp} is
\begin{equation}
\label{var-scaling} 
\mathbb{V}_k(N) = N^{4/3}\mathcal{V}(\kappa)
\end{equation}
where $\mathcal{V}((\kappa)\simeq C_2\kappa$ when $\kappa\to 0$ as we stated in Eq.~\eqref{var-extreme}. The asymptotically exact behavior in the complimentary $\kappa\to \infty$ limit stated in Eq.~\eqref{var-extreme} is established in Sec.~\ref{sec:EA}. 

Simulation results \cite{Ginestra} support an algebraic decay in the $\kappa\to \infty$ limit: $\mathcal{V}\sim \kappa^{-\gamma}$. The uncertainty \cite{Ginestra} in the magnitude of the exponent is significant: $\gamma=0.75\pm 0.15$. Our theoretical prediction for this exponent is $\gamma=\frac{1}{2}$, and we have also derived the amplitude: $\mathcal{V}\simeq \sqrt{2/\kappa}$ as stated in Eq.~\eqref{var-extreme}. 

\section{Outbreak Size Distribution: Exact treatment}
\label{sec:EA}

The critical SIR process admits an exact treatment. Denote by $s, i$ and $r$ the population sizes in the susceptible, infected and recovered individuals. The entire size of the  population is $N$, so 
\begin{equation}
\label{all}
s + i + r = N
\end{equation}

Due to the constraint \eqref{all}, the state of the process can be described by any pair of variables $s, i, r$. We choose $(i,x)$ with $x=N-s$. For the critical SIR process, in the interesting regime $s$ is close to $N$, viz.  $N-s\ll N$, and hence $x$ is a more convenient variable than $s$. The constraint \eqref{all} shows that $x=i+r$, so $x\geq i$. 

Infection and recovery events are symbolically  
\begin{subequations}
\begin{align}
\label{inf}
& (i,x) \to (i+1,x+1) ~\quad \text{rate} ~~ i(N-x)/N\\
\label{rec}
& (i,x) \to (i-1,x) \quad\qquad \text{rate} ~~ i
\end{align}
\end{subequations}

Denote by $t(i,x)$ the number of transitions from the state $(i,x)$ to termination. We are mostly interested in $t(k,k)$, i.e., starting with $k$ infected and no recovered. The process terminates at some state $(0,n)$, where $n$ is the size of the outbreak. The rules \eqref{inf} and \eqref{rec} show that the quantity $i-2x$ decreases by 1 in each transition. Thus starting at $(i,x)=(k,k)$ gives $i-2x=-k-T$ after $T$ transitions, and in particular 
\begin{equation}
\label{n:tkk}
n=\frac{k+t(k,k)}{2}
\end{equation}

The rates of the processes  \eqref{inf} and \eqref{rec} imply that they occur with probabilities
\begin{equation}
p_+(x)=\frac{N-x}{2N-x}\,, \quad p_-(x)=\frac{N}{2N-x}
\end{equation}
The stochastic transition time $t(i,x)$ evolves according to the rules
\begin{equation}
\label{tix:rules}
t(i,x) =
\begin{cases}
1+t(i+1,x+1) & \text{prob} ~p_+(x)\\
1+t(i-1,x)      & \text{prob} ~p_-(x)
\end{cases}
\end{equation}

\subsection{Average number of transitions: Exact solution}

Averaging \eqref{tix:rules} we find that $T_1(i,x)=\langle t(i,x)\rangle$ satisfies
\begin{equation}
\label{T1:eq}
T_1 = 1 + p_+T_1(i+1,x+1)+ p_-T_1(i-1,x) 
\end{equation}
To avoid cluttering of formulae, we write $p_\pm \equiv p_\pm(x)$ and $T_1\equiv T_1(i,x)$  when there is no confusion. 
The recurrence \eqref{T1:eq} should be solved subject to the boundary condition 
\begin{equation}
\label{BC:cat}
T_1(0,x)=0
\end{equation}

The boundary-value problem \eqref{T1:eq}--\eqref{BC:cat} admits an exact solution
\begin{eqnarray}
\label{T1:exact}
T_1 &=& i +2(N-x)\nonumber\\
& - & \sum_{j=1}^{N-x} \left(\frac{N}{N+j}\right)^{i+N-x-j} B_{j}^{(N-x)}(N)
\end{eqnarray}
with $B_{j}^{p}(N)$ determined recurrently from
\begin{equation}
\begin{split}
\label{Bjp}
& B_{j}^{(p)}(N)  = \frac{p}{p-j}\,B_{j}^{(p-1)}(N) \,, \quad j=1,\ldots,p-1\\
& B_{p}^{(p)}(N)  = 2p - \sum_{j=1}^{p-1} \left(\frac{N}{N+j}\right)^{p-j} B_{j}^{(p)}(N) 
\end{split}
\end{equation}
One can verify by direct substitution that Eq.~\eqref{T1:exact} with amplitudes determined from the recurrent relations \eqref{Bjp} satisfies \eqref{T1:eq}--\eqref{BC:cat}. In Appendix \ref{ap:exact}, we describe the rationale behind the ansatz \eqref{T1:exact}. 

Specializing Eq.~\eqref{T1:exact} to $i=x=k$ we obtain
\begin{eqnarray}
\label{Tkk}
T_1(k,k) &=& 2N -k \nonumber\\
& - & \sum_{j=1}^{N-k} \left(\frac{N}{N+j}\right)^{N-j} B_{j}^{(N-k)}(N)
\end{eqnarray}
The average size of an outbreak, $\mathbb{E}_k(N)=[k+T_1(k,k)]/2$ as it follows from Eq.~\eqref{n:tkk}, is therefore 
\begin{equation}
\label{av:outbreak}
\mathbb{E}_k(N)=N-\frac{1}{2}\sum_{j=1}^{N-k} \left(\frac{N}{N+j}\right)^{N-j} B_{j}^{(N-k)}(N)
\end{equation}

An exact formula \eqref{av:outbreak} for the average size of the outbreak valid for arbitrary $k$ and $N$ was a pleasant outcome, yet we have not succeeded so far in extracting an asymptotic behavior of the sum in Eq.~\eqref{av:outbreak}. There are two technical challenges: (i) the amplitudes determined by the recurrence relations \eqref{Bjp}, are unwieldy; (ii) the sum in Eq.~\eqref{av:outbreak} involves $N-k$ terms. One can find compact exact results when $N-k=O(1)$, see Appendix \ref{ap:exact}. Since $k=O(N^{1/3})$ in the interesting range, the number of terms in the sum in Eq.~\eqref{av:outbreak} is close to $N\gg 1$. 

One still hopes to extract the currently unknown scaled size distribution $\mathcal{E}(\kappa)$ from the exact solution \eqref{Bjp}--\eqref{av:outbreak}, and perhaps even sub-leading corrections to the leading asymptotic behavior, Eq.~\eqref{av-scaling}. A numerical integration of \eqref{Bjp}--\eqref{av:outbreak} may prove advantageous to simulations since one does not need to perform the averaging. Here we merely demonstrate how to obtain explicit analytical results for sufficiently small $N$. Solving the recurrence in the top line Eq.~\eqref{Bjp} gives
\begin{equation}
B_{j}^{(p)}  = \binom{p}{j}\,B_{j}, \qquad B_j \equiv B_{j}^{(j)}
\end{equation}
and therefore the bottom in Eq.~\eqref{Bjp} turns into a system of linear equations 
\begin{equation}
\label{B:eq}
\sum_{j=1}^{p-1} \left(\frac{N}{N+j}\right)^{p-j} \binom{p}{j}\,B_{j} + B_p = 2p
\end{equation}
The average size of an outbreak simplifies to 
\begin{equation}
\label{av:out-simple}
\mathbb{E}_k=N-\frac{1}{2}\sum_{j=1}^{N-k} \left(\frac{N}{N+j}\right)^{N-j} \binom{N-k}{j} B_j
\end{equation}
with $B_j=B_j(N)$ found by solving Eqs.~\eqref{B:eq}. 

\begin{table}[t]
\begin{center}
\label{Tab:E12}
\begin{tabular}{|l|l|l|l|l}
\hline
$N$ &$\mathbb{E}_1(N)$ &$\mathbb{E}_2(N)$ \\
\hline
$2$ &   $\tfrac{4}{3}$                                                      & $2$ \\
\hline
$3$ &   $\tfrac{63}{40}$                                                  & $\tfrac{39}{16}$ \\
\hline
$4$ &   $\tfrac{664}{375}$                                              &  $\tfrac{3148}{1125}$\\
\hline
$5$ &   $ \tfrac{10336175}{5334336}$                           & $\tfrac{7372655}{2370816}$ \\
\hline
$6$ &    $\tfrac{1110011389}{532445760}$                    & $\tfrac{819586429}{242020800}$ \\
\hline
$7$ &    $\frac{12370766783106977}{5579979964416000}$     & $\frac{9366162186882977}{2575375368192000}$ \\
\hline
$8$ &    $\frac{61815870564834991712}{26443021298445984375}$     & $\frac{47712925872330466712}{12340076605941459375}$ \\
\hline
$9$ &    $\frac{1322599539600214934018051151}{540030848151585190400000000}$     & $\frac{1036700855284669833218051151}{254132163836040089600000000}$ \\
\hline
\end{tabular}
\caption{The average size of an outbreak starting with one infected individual, $\mathbb{E}_1(N)$, and with two  infected individuals, $\mathbb{E}_2(N)$, for $N=2,\ldots,9$.}  
\end{center}
\end{table}

\begin{figure}
\centering
\includegraphics[width=7.89cm]{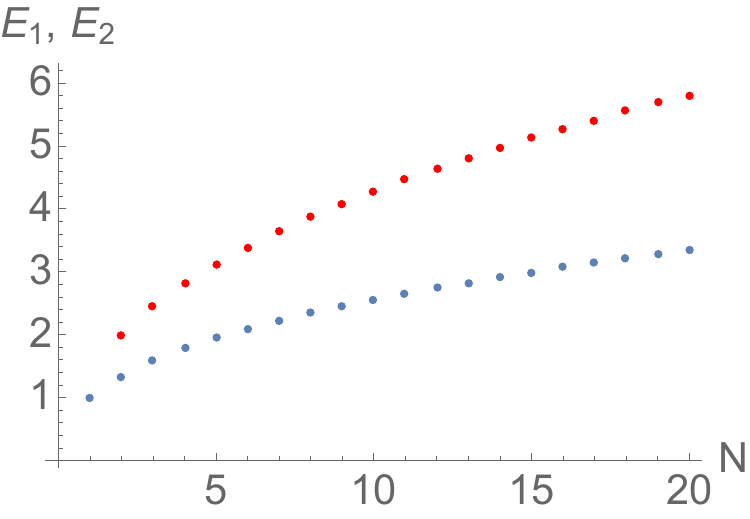}
\caption{Bottom: The average size of an outbreak for epidemics starting with a single infected individual, $\mathbb{E}_1(N)$.
Top: The average size of an outbreak for epidemics starting with two infected individuals, $\mathbb{E}_2(N)$.}
\label{Fig:E12}
\end{figure}

Exact results for $\mathbb{E}_1(N)$ and $\mathbb{E}_2(N)$ are given in Table~I for $N\leq 9$, and plotted in Fig.~\ref{Fig:E12} for $N\leq 20$. Even for such small populations, the exact results are rather close to the asymptotic behavior \eqref{NkN-asymp}.

\subsection{Average number of transitions: Continuum treatment}

An often potent line of attack on asymptotic behavior of solution to discrete problems relies on employing continuum methods. In the present case we assume that $T_1(i,x)$ is a smooth function of $i$ and $x$, and we expand  $T_1(i+1,x+1)$ and $T_1(i-1,x)$ appearing in Eq.~\eqref{T1:eq} up to the second order
\begin{equation}
\label{T1:exp}
\begin{split}
T_1(i-1,x) &= T_1 - \partial_i T_1 + \tfrac{1}{2}\partial_i^2 T_1  \\
T_1(i+1,x+1) &= T_1 + \partial_i T_1 + \partial_x T_1 \\
&+ \tfrac{1}{2}\partial_i^2 T_1 + \tfrac{1}{2}\partial_x^2 T_1 +  \partial_i \partial_x T_1
\end{split}
\end{equation}
Here we use shorthand notation $\partial_i=\partial/\partial i$, $\partial_x=\partial/\partial x$, etc. for the partial derivatives. Inserting \eqref{T1:exp} into \eqref{T1:eq} and keeping only dominant terms we obtain
\begin{equation}
\label{t:ix}
\partial_i^2 T_1 + \partial_x T_1 + 2 =   \frac{x}{N}\,\partial_i T_1
\end{equation}
Suppose that the scaling in the interesting range is
\begin{equation}
\label{abc}
i\sim N^\alpha, \quad x\sim N^\beta, \quad T_1\sim N^\gamma
\end{equation}
Plugging \eqref{abc} into \eqref{t:ix} we find that the terms in  are comparable only when $\alpha=\frac{1}{3}$ and $\beta=\gamma=\frac{2}{3}$. Thus we re-scale the variables 
\begin{equation}
\label{IX}
i = N^{1/3} I, \quad x = N^{2/3} X
\end{equation}
and the average number of transitions 
\begin{equation}
\label{TIX:def}
T_1(i,x) = N^{2/3} \mathcal{T}(I, X)
\end{equation}
One can verify that terms not included in \eqref{t:ix} are subdominant. For instance, computing the second derivates gives
$\partial_i \partial_x T_1=O(N^{-1/3})$ and $\partial_x^2 T_1=O(N^{-2/3})$, so these derivatives can indeed be dropped. 

The transformation \eqref{IX}--\eqref{TIX:def} turns \eqref{t:ix} into a partial differential equation (PDE)
\begin{equation}
\label{TIX}
\frac{\partial^2 \mathcal{T}}{\partial I^2} + \frac{\partial \mathcal{T}}{\partial X} + 2= X\frac{\partial \mathcal{T}}{\partial I} 
\end{equation}
for the re-scaled transition time $\mathcal{T}(I, X)$. 

We must solve Eq.~\eqref{TIX} in the quadrant $I\geq 0$ and $X\geq 0$. The boundary condition, Eq.~\eqref{BC:cat}, yields 
\begin{equation}
\label{BC:cont}
\mathcal{T}(0, X) = 0
\end{equation}

Solving the boundary-value problem \eqref{TIX}--\eqref{BC:cont} is an intriguing challenge that we leave for the future. Here we limit ourselves by a simpler problem of computing the asymptotic behavior of $T_1(k,k)$ when $k\gg N^{1/3}$. 

In the realm of the framework \eqref{IX}--\eqref{BC:cont}, we should learn how to extract  the large $I$ behavior of $\mathcal{T}(I, X)$. This can be done by noting that when $I\gg 1$, the  diffusion term can be dropped from Eq.~\eqref{TIX}. Thus we arrive at the first order PDE
\begin{equation}
\label{PDE}
\frac{\partial \mathcal{T}}{\partial I}  - \frac{1}{X}\,\frac{\partial \mathcal{T}}{\partial X} = \frac{2}{X}
\end{equation}
Introducing new variables
\begin{equation}
\label{uv}
u = I + \tfrac{1}{2}X^2\,, \qquad v = I - \tfrac{1}{2}X^2
\end{equation}
we recast \eqref{PDE} into 
\begin{equation}
\label{PDE:uv}
\frac{\partial \mathcal{T}}{\partial v}  = \frac{1}{\sqrt{u-v}}
\end{equation}
The solution is
\begin{equation*}
\mathcal{T}=-2\sqrt{u-v}+f(u)
\end{equation*}
with an arbitrary function $f(u)$. The boundary condition \eqref{BC:cont} gives $\mathcal{T} = 0$ when $v=-u$. This fixes $f(u)=2\sqrt{2u}$. Combining $\mathcal{T}= 2\sqrt{2u}-2\sqrt{u-v}$ and \eqref{uv} we obtain
\begin{equation}
\label{far}
\mathcal{T}(I, X) = 2\sqrt{2I+X^2}-2X
\end{equation}

We want to determine $T_1(k,k) = N^{2/3} \mathcal{T}(\kappa, N^{-1/3}\kappa)$. Thus $I=\kappa\gg 1$ and $X=0$ as we always consider large populations, $N\gg 1$. More precisely, setting $X=0$ amounts for a tacit assumption $\kappa\ll N^{1/3}$. Summarizing, our asymptotic results are valid in the range 
\begin{equation}
\label{k-bounds}
N^{1/3}\ll k \ll N^{2/3}
\end{equation}
The upper and lower bounds are well separated when $N^{1/3}\gg 1$. This is satisfied for large populations, yet the convergence may be slow as the effective small parameter is $N^{-1/3}$. 

Thus $T_1(k,k) = N^{2/3} \mathcal{T}(\kappa, 0)$ when the bounds \eqref{k-bounds} are obeyed. Using Eq.~\eqref{far} we get $T_1(k,k) = \sqrt{8kN}$ which we insert into $\mathbb{E}_k(N)=[k+T_1(k,k)]/2$ obtained after averaging Eq.~\eqref{n:tkk}.  Keeping only the leading term gives $\mathbb{E}_k(N) = \sqrt{2kN}$. This completes the derivation of the large $\kappa$ behavior announced in Eq.~\eqref{av-extreme}. 

\subsection{Variance}

To derive the governing equations for the variance we first take the square of Eq.~\eqref{tix:rules}. Performing averaging we find that 
$T_2(i,x)=\langle t^2(i,x)\rangle$ satisfies 
\begin{eqnarray}
\label{T2:eq}
T_2 &=& 1 + p_+T_2(i+1,x+1)+ p_- T_2(i-1,x) \nonumber\\
&+& 2 p_+ T_1(i+1,x+1)+  2p_- T_1(i-1,x)
\end{eqnarray}
Again we shortly write $p_\pm \equiv p_\pm(x)$ and $T_2\equiv T_2(i,x)$. 

We now subtract the square of Eq.~\eqref{T1:eq} from Eq.~\eqref{T2:eq}  and find that the variance 
\begin{equation}
V(i,x) = \langle t^2(i,x)\rangle - \langle t(i,x)\rangle^2
\end{equation}
satisfies
\begin{eqnarray}
\label{V:eq}
V(i,x)  &=&p_+V(i+1,x+1)+ p_- V(i-1,x) \nonumber\\
&+& p_+ p_- [T_1(i+1,x+1) - T_1(i-1,x)]^2
\end{eqnarray}
We do not attempt to solve \eqref{V:eq} and switch to the continuum treatment. Similar to \eqref{T1:exp} we expand the variance 
\begin{equation}
\label{V:exp}
\begin{split}
V(i-1,x) &= V - \partial_i V + \tfrac{1}{2}\partial_i^2 V  \\
V(i+1,x+1) &= V + \partial_i V + \partial_x V \\
&+ \tfrac{1}{2}\partial_i^2 V + \tfrac{1}{2}\partial_x^2 V +  \partial_i \partial_x V
\end{split}
\end{equation}
Plugging the expansions \eqref{T1:exp} and \eqref{V:exp} into \eqref{V:eq} and keeping only dominant terms we obtain
\begin{equation}
\label{V:ix}
\partial_i^2 V + \partial_x V -\frac{x}{N}\, \partial_i V +  2(\partial_i T_1)^2 = 0
\end{equation}
We use the same rescaled variables \eqref{IX} as before, and seek the variance in the scaling form
\begin{equation}
\label{VIX:def}
V(i,x) = N^{4/3} \mathcal{V}(I, X)
\end{equation}
The transformation \eqref{IX} and \eqref{VIX:def} turns \eqref{V:ix} into
\begin{equation}
\label{VIX}
\frac{\partial^2 \mathcal{V}}{\partial I^2} + 2 \left(\frac{\partial \mathcal{T}}{\partial I}\right)^2= X\frac{\partial \mathcal{V}}{\partial I} 
- \frac{\partial \mathcal{V}}{\partial X}
\end{equation}

When $I\gg 1$, we can drop again the diffusion term from \eqref{VIX}.  We also use the asymptotic expression \eqref{far} for $\mathcal{T}(I,X)$ and arrive at the first order PDE
\begin{equation}
\label{VIX:1}
X\frac{\partial \mathcal{V}}{\partial I} - \frac{\partial \mathcal{V}}{\partial X} = \frac{8}{2I+X^2}
\end{equation}
Using the variables \eqref{uv} we recast \eqref{VIX:1} into
\begin{equation}
\label{PDE:V}
\frac{\partial \mathcal{V}}{\partial v}  = \frac{2}{u\sqrt{u-v}}
\end{equation}
which is integrated to give $\mathcal{V}=4u^{-1}\left[\sqrt{2u}-\sqrt{u-v}\right]$, or
\begin{equation}
\label{far:V}
\mathcal{V}(I, X) = 8\frac{\sqrt{2I+X^2}-X}{2I+X^2}
\end{equation}

Setting $I=\kappa$ and $X=0$ in Eq.~\eqref{far:V} gives the large $\kappa$ behavior: $\mathcal{V}(\kappa, 0) = 4 \sqrt{2/\kappa}$. Using Eq.~\eqref{n:tkk} we find $\mathbb{V}_k(N)=N^{4/3}\,\tfrac{1}{4}\mathcal{V}(\kappa, 0)$, viz. the large $\kappa$ behavior announced in Eq.~\eqref{var-extreme}.

\section{Duration of Outbreaks}
\label{sec:SIR-time}

Some basic features of the duration of the outbreaks in the critical SIR process in a finite system can be extracted from the temporal behaviors in the infinite-population limit. We first recall these infinite-population results in the simplest case with one initially infected individual \cite{Bailey,bk12}. The probability $P_i(t)$ to have $i$ infected individuals at time $t$ satisfies 
\begin{subequations}
\begin{align}
\label{Pi:eq}
\dot P_i &  = (i-1)P_{i-1}-2iP_i+(i+1)P_{i+1}, \quad i\geq 1\\
\label{P0:eq}
\dot P_0 & = P_1
\end{align}
\end{subequations}
where dot denotes the derivative with respect to time. 

A solution of an infinite set of equations \eqref{Pi:eq}--\eqref{P0:eq} subject to the initial condition $P_i(0)=\delta_{i,1}$ reads 
\begin{subequations}
\begin{align}
\label{Pi:sol}
P_i(t) &=(1+t)^{-2}\, \tau^{i-1}, \quad i\geq 1\\
\label{P0:sol}
P_0(t)&= \tau \equiv \frac{t}{1+t}
\end{align}
\end{subequations}
This soultion can be verified by a direct substitution, or derived using e.g. generating function techniques [see Appendix \ref{ap:time} for details]. 
The probability that the outbreak is still alive at time $t$, is 
\begin{equation}
\label{s}
P(t)=\sum_{i\geq 1} P_i(t)=1-P_0(t)=\frac{1}{1+t}
\end{equation}
The average number of infected individuals in outbreaks which are still alive at time $t$ is therefore
\begin{equation}
\label{iav}
\langle i\rangle=\frac{\sum iP_i(t)}{\sum P_i(t)}=1+t
\end{equation}

A general solution of Eqs.~\eqref{Pi:eq}--\eqref{P0:eq} describing an infinite-population limit subject to an arbitrary number of initially infected individuals  is somewhat cumbersome, it is presented in Appendix \ref{ap:time}. 

In a finite population, the infection process eventually comes to an end. To estimate heuristically this final time $t_\text{f}$ one uses \eqref{iav} to express \cite{bk} the final size of the outbreak through the final time: $n_\text{f} \sim\int^{t_\text{f}} dt\,\langle i\rangle \sim t_\text{f}^2$.  The maximal outbreak size scales as $n_*\sim N^{2/3}$ (see e.g. \cite{bk,KS,bk12}) and hence the maximal duration is
\begin{equation}
\label{TN3}
t_* \sim n_*^{1/2} \sim N^{1/3}
\end{equation}

The average duration of the outbreak is formally
\begin{equation}
\label{time-av}
 \mathbb{E}[t]  = \int_0^\infty dt\,t\left(-\frac{dP}{dt}\right) = \int_0^\infty dt\,t P_1(t)
\end{equation}
Recalling that $P_1(t)=(1+t)^{-2}$ in the infinite-population limit (equivalently, for the critical branching process), we notice that the integral in \eqref{time-av} diverges. We should use, however, the finite upper limit given by \eqref{TN3}. This leads to an estimate
\begin{equation}
\label{time-av-log} 
\mathbb{E}_1[t]   \simeq    \int_0^{\sqrt[3]{N}} dt\,\frac{t}{(1+t)^2}  \simeq \frac{1}{3}\,\ln N
\end{equation}
where the subscript reminds that the process begins with a single infected individual. The logarithmic growth of the average duration time was predicted by Ridler-Rowe many years ago \cite{rr}, albeit with incorrect amplitude; the correct amplitude $1/3$ is easy to appreciate \cite{caveat}, it was argued and numerically supported in \cite{bk,bk12}. The above argument also suggests the more precise asymptotic
\begin{equation}
\label{time-av-log-1} 
\mathbb{E}_1[t]  = \tfrac{1}{3}\ln N + c_1 + o(1/N)
\end{equation}
Since the logarithm is a slowly growing function, the sub-leading constant term $c_1$ significantly contributes to the average duration. The computation of the sub-leading term requires much more comprehensive analysis than what we have used so far.

If the number of initially infected individuals is sufficiently small, $k\ll N^{1/3}$, one can generalize the prediction \eqref{time-av-log} for the average duration of an outbreak without using a complete solution of the infinite-population limit (Appendix \ref{ap:time}), it suffices to rely on the independence of infection processes generated by each initially infected individual. The probability that the infection is over at time $t$ is $P_0^k$, with $P_0$ given by \eqref{P0:sol}. Thus $-dP_0^k/dt=kP_0^{k-1}/(1+t)^2$ is the probability density that the infection is eradicated at time $t$, from which
\begin{equation}
\label{time-av-log-k} 
\mathbb{E}_k[t]   \simeq \int_0^{\sqrt[3]{N}} dt\,\frac{kt^k}{(1+t)^{k+1}}  \simeq \frac{k}{3}\,\ln N
\end{equation}
implying that the average duration of the outbreak exhibits a simple logarithmic scaling with amplitude proportional to the initial number of infected individuals. 

Equation \eqref{time-av-log-k} suggests a plausible scaling form 
\begin{equation}
\label{av-time-scaling}
\mathbb{E}_k[t] = \frac{N^{1/3} \ln N}{3}\,\, \Theta(\kappa)
\end{equation}
with $\Theta(\kappa)\simeq \kappa$ when $\kappa\to 0$ to match the asymptotic \eqref{time-av-log-k}. The scaled size of initially infected individuals is again $\kappa=k/N^{1/3}$. The $N-$dependent pre-factor, however, contains an additional logarithmic factor. The unexpected logarithmic factor was originally observed in simulations \cite{Ginestra}. In Appendix \ref{ap:det}, we probe the large $\kappa$ using a deterministic approach. This analysis supports the scaling form \eqref{av-time-scaling}. Overall, the extremal behaviors of the scaled average time are
\begin{equation}
\label{av-time-extreme}
\Theta(\kappa) =
\begin{cases}
\kappa               &\text{when}\quad \kappa\ll 1\\
\sqrt{2/\kappa}   &\text{when}\quad \kappa\gg 1
\end{cases}
\end{equation}

The variance can be probed similarly to the average, see \eqref{time-av-log}. One establishes the scaling law
\begin{equation}
\label{time-var} 
\mathbb{V}_1[t]   \sim    \int_0^{\sqrt[3]{N}} dt\,\frac{t^2}{(1+t)^2}  \sim N^{1/3}
\end{equation}
but not an amplitude. Independence gives $\mathbb{V}_1[t] \sim kN^{1/3}$ when $k\ll N^{1/3}$, and hence the hypothetical scaling form of the variance is
\begin{equation}
\label{var-time-scaling}
\mathbb{V}_k[t] = N^{2/3}\, \Theta_2(\kappa)
\end{equation}
with  $\Theta_2(\kappa)\sim \kappa$ when $\kappa\to 0$. In contrast to the average, the scaled form for the variance does not contain a logarithmic factor. 

Simulations \cite{Ginestra} support the scaling law \eqref{var-time-scaling} and the linear small $\kappa$ behavior. Simulations also suggest \cite{Ginestra} that the scaled distribution $\Theta_2(\kappa)$ is inversely proportional to $\kappa$ in the large $\kappa$ limit. Thus
\begin{equation}
\label{var-time-extreme}
\Theta_2(\kappa)\sim
\begin{cases}
\kappa          & \kappa\to 0\\
\kappa^{-1}   & \kappa\to\infty
\end{cases}
\end{equation}

%The scaling behavior of the average outbreak duration seems rather tricky. A scaling form with pre-factor being a product of an algebraic and a logarithmic in $N$ terms has been used in \cite{Ginestra} to fit simulation data. This scaling form is notably different from three other scaling forms that all have a standard purely algebraic in $N$ pre-factor [cf. Eqs.~\eqref{av-scaling}, \eqref{var-scaling}, and \eqref{var-time-scaling}]. It may be worthwhile to try to fit numerical data for the average outbreak duration with the standard scaling form \eqref{av-time-scaling}, but allowing the possibility of an anomalous small $\kappa$ behavior such as $\Theta(\kappa)\simeq -\kappa \ln \kappa$  argued above. 

\section{Conclusions}

We have studied the critical SIR process starting with a large number of initially infected individuals, $k\gg 1$. Particularly interesting behaviors emerge when $k$ scales as a cubic root of the population size, $k\sim N^{1/3}$. The critical SIR process exhibits large fluctuations, so we have relied on the stochastic formulation. We have treated the problem using a combination of  exact calculations and asymptotic methods. We have focused on the size and duration of the outbreaks, more precisely on the average and variance of these quantities. The analysis of the size of outbreaks is rather detailed, Secs.~\ref{sec:SIR}--\ref{sec:EA}. 

Our analytical and asymptotic predictions qualitatively agree with simulation results \cite{Ginestra}. Whenever there is a discrepancy between simulation results and theoretical predictions, it seems that data may be fitted using slightly different scaling forms, viz. simpler than the fits used in Ref.~\cite{Ginestra}. The chief reason for subtle behaviors are algebraic tails, the average size and duration of outbreaks are especially sensitive to these tails. 

We have found an exact expression for the average size of outbreaks valid for arbitrary $N$ and $k$. Extracting the scaled size distribution $\mathcal{E}(\kappa)$ and sub-leading corrections to the leading behavior \eqref{av-scaling} is left for future work. The derivation of the exact average size in Sec.~\ref{sec:EA} can be probably generalized to establish the variance and cumulants (perhaps even the cumulant generating function). Continuum methods allow, in principle, to determine the scaling functions, one should be able to solve linear PDEs with non-constant coefficients. 

%We have used continuum methods in extracting asymptotic behaviors of the scaling functions. 

\vskip 1cm
\noindent 
{\bf Acknowledgments.} I want to thank Ginestra Bianconi and Sid Redner for collaboration on similar problems, G. Bianconi and F. Radicchi for sending a preliminary version of Ref.~\cite{Ginestra}, and Boston University  Network group for discussions and encouragement.  I am also grateful to the referee who has given a simple derivation of Eq.~\eqref{Ank-gamma} presented in Appendix~\ref{ap:record}.

\appendix

\section{Derivation of Eq.~\eqref{Ank-gamma}}
\label{ap:record}

Plugging \eqref{An_sol} into the sum in Eq.~\eqref{Ank-sum}, one reduces the sum to a hypergeometric series \cite{Knuth}. Using identities involving hypergeometric series \cite{Knuth}, as well as identities involving the gamma function (particularly, the duplication formula), one computes the sum in \eqref{Ank-sum} and arrives at the announced expression \eqref{Ank-gamma} for $A_n^{(k)}$. This derivation relies on identities for the hypergeometric series that are little known. The intermediate result, Eq.~\eqref{Ank-sum}, also requires a rather technical derivation. 

We now present an alternative derivation of Eq.~\eqref{Ank-gamma} directly from the generating function \eqref{gen-k}. Employing the Cauchy theorem, reduces the problem to computing the integral 
\begin{equation}
\label{Ank-Cauchy}
A_n^{(k)} = \oint_{C_0} \frac{dz}{2\pi \ii}\,\frac{ \left\{1 - \sqrt{1-z}\right\}^k}{z^{n+1}}
\end{equation} 
over a small simple counter-clockwise contour $C_0$  around the origin in the complex $z$ plane. To simplify the computation of the integral, let us try to change the variable $z$ so that the square root $\sqrt{1-z}$ would become a linear function of a novel complex variable. The change of variable, $1-z=(2v-1)^2$, achieves this goal and reduces the integral in \eqref{Ank-Cauchy} to 
\begin{equation}
\label{Ank-Cauchy-v}
A_n^{(k)} =\oint_{C_1} \frac{dv}{2\pi \ii}\,\frac{(1-v)^{-(n+1-k)}}{2^{2n-k-1}} \left[\frac{1}{v^n}-\frac{2}{v^{n+1}}\right]
\end{equation}
over the appropriate contour $C_1$. The integral in \eqref{Ank-Cauchy-v} is found without a calculation, it suffices to use the residue theorem  and the binomial theorem \cite{Knuth}
\begin{equation}
(1-v)^{-m}=\sum_{\ell\geq 0} \frac{\Gamma(m+\ell)}{\Gamma(m)\, \Gamma(\ell+1)}\,v^\ell
\end{equation}
The final outcome is Eq.~\eqref{Ank-gamma}. 

The quantity $A_n^{(k)}$ admits an interesting interpretation in terms of records of a one-dimensional random walk discrete in time and with an arbitrary symmetric and continuous jump distribution \cite{Satya}. Namely, $A_n^{(k)}$ is the probability that such a random walk has $k$ maxima in $n$ time steps, with the last position being the maximum. Formulas similar to \eqref{gen-k} and \eqref{Ank-gamma} appear in the context of universal record statistics of random walks \cite{Satya}. 

It would be interesting to extend the relation between the outbreak size distribution and the records of a one-dimensional discrete in time random walk to more complicated versions of the critical branching process, such as the critical branching process with triplication rather than duplication or the discrete time critical branching process.

\section{Deterministic treatment}
\label{ap:det}

For the SIR process starting with a single infected individual, the deterministic framework is applicable only in the super-critical regime if the process enters the run-away regime. In this situation, the epidemic ends only after a finite fraction of the population [see Eq.~\eqref{rR}] catches the disease, and this fraction can be determined using the deterministic framework. If the initial number of infected individuals is very large, the deterministic framework always applies and one can describe the SIR process via the system of differential equations for the densities $S(t)$ of susceptible, $I(t)$ of infected, and $R(t)$ of recovered individuals. 

For the critical SIR process, $R_0=1$, the deterministic rate equations read
\begin{subequations}
\begin{align}
\label{S:eq}
\dot S &  = -S I\\
\label{I:eq}
\dot I& = -I + S I\\
\label{R:eq}
\dot R& = I
\end{align}
\end{subequations}
Equations \eqref{S:eq}--\eqref{R:eq} are consistent with the conservation law $S(t)+I(t)+R(t)=1$. The initial condition 
\begin{equation}
\label{IC:SIR}
S(0)=1-\epsilon, \quad I(0)=\epsilon, \quad R(0)=0
\end{equation}
Below we always assume that
\begin{equation}
N^{-1}\ll \epsilon\ll 1
\end{equation}
In this range, the number of initially infected individuals, $\epsilon N$, is large and the deterministic framework should be asymptotically correct. Setting $\epsilon\ll 1$ is not necessary, but this assumption implies that the initial fraction of infected individuals is small; this is natural since we assume that the containment measures making the process critical have started when still a microscopic fraction of the population caught the disease. Furthermore, assuming that $\epsilon\ll 1$ allows one to derive much more explicit (asymptotic) formulas than in the general case. 

To solve \eqref{S:eq}--\eqref{R:eq} subject to \eqref{IC:SIR} it is convenient to treat $S$ as a time variable. Dividing \eqref{I:eq} by \eqref{S:eq} and integrating gives 
\begin{equation}
\label{I:sol}
I = 1-S+\ln\frac{S}{1-\epsilon}
\end{equation}
Similarly dividing \eqref{R:eq} by \eqref{S:eq} and integrating yields
\begin{equation}
\label{R:sol}
R = -\ln\frac{S}{1-\epsilon}
\end{equation}
The outbreak ends when $I_\text{f}=0$. Equation \eqref{I:sol} shows that the final fraction of susceptible is implicitly determined by 
\begin{equation}
\label{Sf:exact}
1-S_\text{f} = -\ln\frac{S_\text{f} }{1-\epsilon}
\end{equation}
from which 
\begin{equation}
\label{Rf}
R_\text{f} = 1-S_\text{f} \simeq  \sqrt{2\epsilon}
\end{equation}
when $\epsilon\ll 1$. Since $R_\text{f}=N^{-1}\mathbb{E}_k(N)$ and $\epsilon=k/N$, we can re-write \eqref{Rf} as $\mathbb{E}_k(N)=\sqrt{2kN}$, leading to the asymptotic $\Psi(\kappa) =\sqrt{2\kappa}$ when $\kappa\gg 1$. 

The applicability of the deterministic framework for sufficiently large $k$ is intuitively clear. We have determined the crossover $k\sim N^{1/3}$ when the stochastic effects become important using the stochastic framework. It would be interesting to deduce the crossover in the realm of the deterministic framework. 

The applicability of the deterministic framework on the final stage with a few remaining infected individuals is questionable. We know, however, that for the critical SIR process starting from a single infected individual, large outbreaks are of the order of $N^{2/3}$. Thus when $\kappa\gg 1$, the outbreak size is $N^{2/3}\sqrt{2\kappa} + O(N^{2/3})$. The first term describes the deterministic stage, while the second term accounts for the stochastic final stage. The deterministic first term dominates when  $\kappa\gg 1$. Therefore in this regime, one can use the deterministic framework. 

To estimate the average duration time using the deterministic framework, we combine \eqref{S:eq} and \eqref{I:sol} and obtain
\begin{equation}
\label{av-t-int}
\mathbb{E}[t] = \int_{S_\text{f}}^{S(0)}\frac{d\sigma}{\sigma\left[1-\sigma+\ln\frac{\sigma}{1-\epsilon}\right]}
\end{equation}
Writing $\sigma=1-\sqrt{2\epsilon}\,y$, expanding the denominator in the integral in Eq.~\eqref{av-t-int} in powers of $\epsilon\ll 1$, and recalling that 
$1-S(0)=\epsilon$ and $1-S_\text{f} \simeq  \sqrt{2\epsilon}$, we obtain  
\begin{eqnarray}
\label{av-t-asymp}
\mathbb{E}_k[t] &\simeq & \sqrt{\frac{2}{\epsilon}}\int_0^1 \frac{dy}{1+\frac{\epsilon}{2}-y^2}  \nonumber \\
                         &\simeq & \frac{\ln(8/\epsilon)}{\sqrt{2\epsilon}} = N^{1/3}\, \frac{\ln(8N^{2/3}/\kappa)}{\sqrt{2\kappa}}  \nonumber \\
                         &\simeq & \frac{N^{1/3}\,\ln N}{3}\,\,\sqrt{\frac{2}{\kappa}}
\end{eqnarray}
(We have taken into account that $\epsilon=k/N=\kappa/N^{2/3}$.) This asymptotic is consistent with the scaling form \eqref{av-time-scaling} and gives the announced large $\kappa$ asymptotic in Eq.~\eqref{av-time-extreme}.

\section{Exact results}
\label{ap:exact}

In Sec.~\ref{sec:EA}, we have used asymptotic methods to determine the leading behaviors. It is possible to derive exact results for the average outbreak size. Unfortunately, these results are expressed through the sum of a large number of variables with complicated individual terms. Here we explain how one could guess these results; the verification of the guess \eqref{T1:exact} is straightforward. We have made the guess \eqref{T1:exact} by establishing a few explicit exact results. These exact results describe a non-interesting region where the initial number of susceptible individuals is small, so their virtue is that they simplify the guesswork. 

The average number of transitions $T_1(i,x)$ from the state $(i,x)$ to termination satisfies the recurrence \eqref{T1:eq} and the boundary condition \eqref{BC:cat}. We start by solving \eqref{T1:eq}--\eqref{BC:cat} in the simplest cases when $N-x=O(1)$. First, we notice that 
\begin{equation}
\label{hyp}
T_1(i,N)  = i
\end{equation}
Indeed, only recovery events are possible if there are no susceptible. One can also formally derive \eqref{hyp} by specializing \eqref{T1:eq} to $x=N$. This gives $T_1(i,N) = 1 + T_1(i-1,N)$ which in conjunction with $T_1(0,N)=0$ lead to \eqref{hyp}. 

Similarly we specialize Eq.~\eqref{T1:eq} to $x=N-1$ and use Eq.~\eqref{hyp} to obtain 
\begin{equation*}
T_1(i,N-1) = 1+\frac{N}{N+1}\,T_1(i-1,N-1)+\frac{i+1}{N+1}
\end{equation*}
Solving this recurrence subject to $T(0,1)=0$ one obtains
\begin{equation}
\label{hyp:1}
T_1(i,N-1)  = i+2-2\left(\frac{N}{N+1}\right)^i
\end{equation}

Specializing \eqref{T1:eq} to $x=N-2$ and using \eqref{hyp:1} we arrive at the recurrence 
\begin{eqnarray*}
T_1(i,N-2) & = & 1+\frac{N}{N+2}\,T_1(i-1,N-2)\nonumber \\
&+& \frac{2}{N+2}\left[i+3-2\left(\frac{N}{N+1}\right)^{i+1}\right]
\end{eqnarray*}
from which
\begin{eqnarray}
\label{hyp:2}
T_1(i,N-2) & = &i+4-4\left(\frac{N}{N+1}\right)^{i+1} \nonumber \\
& - & \frac{4}{N+1}\left(\frac{N}{N+2}\right)^i
\end{eqnarray}
Similarly we compute
\begin{eqnarray}
\label{hyp:3}
T_1(i,N-3) & = &i+6-6\left(\frac{N}{N+1}\right)^{i+2} \nonumber\\
&-& \frac{12}{N+1}\left(\frac{N}{N+2}\right)^{i+1}\nonumber\\
&-& \frac{12+18N}{(N+1)^2(N+2)}\left(\frac{N}{N+3}\right)^{i}
\end{eqnarray}

We are interested in $T_1(k,k)$. Specializing Eq.~\eqref{hyp} to $i=N$ yields $T_1(N,N) = N$. Similarly 
from \eqref{hyp:1}--\eqref{hyp:3} we extract  
\begin{equation*}
\begin{split}
T_1(N-1,N-1) &= N+1 - 2\left(\frac{N}{N+1}\right)^{N-1} \\
T_1(N-2,N-2) &= N+2 - 4\left(\frac{N}{N+1}\right)^{N-1} \\
                      & -\frac{4}{N+1}\left(\frac{N}{N+2}\right)^{N-2}  \\
T_1(N-3,N-3) &= N+3 - 6\left(\frac{N}{N+1}\right)^{N-1} \\
                      & - \frac{12}{N+1}\left(\frac{N}{N+2}\right)^{N-2} \\
                      &-\frac{12+18N}{(N+1)^2(N+2)}\left(\frac{N}{N+3}\right)^{N-3}
\end{split}
\end{equation*}

Looking at the above expressions for $T_1(N-p,N-p)$ with $p=0,1,2,3$, one guesses the general formula 
\begin{eqnarray}
\label{TNp}
T_1(N-p,N-p) &=& N +p\nonumber\\
& - & \sum_{j=1}^p \left(\frac{N}{N+j}\right)^{N-j} B_{j}^{(p)}
\end{eqnarray}
This is exactly \eqref{Tkk} in different notation. More generally, Eqs.~\eqref{hyp}--\eqref{hyp:3} suggest our chief ansatz, Eq.~\eqref{T1:exact}, which is then straightforwardly verified. 

The amplitudes $B_j^{(p)}(N)$ are rather simple for small $p$, but quickly become cumbersome. Here are a few series of the amplitudes extracted from 
\eqref{Bjp}:
\begin{equation}
\begin{split}
& B_{1}^{(p)}  = 2p \\
& B_{2}^{(p)}  = \frac{2p(p-1)}{N+1} \\
& B_{3}^{(p)}  = \frac{p(p-1)(p-2)(2+3N)}{(N+1)^2(N+2)} 
\end{split}
\end{equation}
Therefore when $p=O(1)$ is fixed 
\begin{eqnarray*}
T_1(N-p,N-p) &=& N + p\left(1-\frac{2}{e}\right) -\frac{2p(p-1) e^{-2}}{N+1}\\
&-& \frac{3p(p-1)(p-2) e^{-3}}{(N+1)^2}+O(N^{-3})
\end{eqnarray*}
when $N\gg 1$. The asymptotic behavior of the average size of the outbreak is therefore
\begin{equation}
\mathbb{E}_{N-p}(N)  = N -\frac{p}{e} -\frac{p(p-1)}{e^2}\,N^{-1}+O(N^{-2})
\end{equation}

\section{The distribution $P_i(t)$}
\label{ap:time}

We want to solve  Eqs.~\eqref{Pi:eq}--\eqref{P0:eq} subject to the initial condition $P_i(t=0) = \delta_{i,k}$. Using the generating function 
\begin{equation}
\label{gen}
g(z,t) = \sum_{i\geq 0} P_i(t) \, z^i
\end{equation}
we reduce the infinite system \eqref{Pi:eq}--\eqref{P0:eq}  of ordinary differential equations to a partial differential equation
\begin{equation}
\label{gen:eq}
\partial_t g = (1-z)^2 \partial_z g
\end{equation}
Introducing the auxiliary variable $\zeta=(1-z)^{-1}$ we recast \eqref{gen:eq} into $\left(\partial_t - \partial_\zeta\right)g = 0$ which is solved to yield
\begin{equation}
\label{gen:zeta}
g(z,t) = G(t+\zeta)
\end{equation}
The function $G$ is fixed by the initial condition
\begin{equation}
\label{G0}
G(\zeta) = g_0(z)
\end{equation}
In terms of the original variable $z$, the solution \eqref{gen:zeta}--\eqref{G0} becomes 
\begin{equation}
\label{gen:sol}
g(z,t) = g_0\!\left(1-\frac{1-z}{1+t - t z}\right)
\end{equation}
This solution is valid for an arbitrary initial condition.  It is useful to rewrite this solution in terms of the reduced generating function 
\begin{equation}
\label{gen-P}
\mathcal{P}(z,t) = \sum_{m\geq 1} P_m(t) \, z^m = g(z,t) - g(0,t) 
\end{equation}
accounting only for the active part. One gets 
\begin{equation}
\label{gen-P:sol}
\mathcal{P}(z,t) = g_0\!\left(1-\frac{1-z}{1+t - t z}\right) - g_0(\tau)
\end{equation}

In the classical case $P_i(0) = \delta_{i,1}$ we get $g_0(z)=z$, so \eqref{gen-P:sol} reduces to 
\begin{eqnarray}
\label{gen-P:1}
\mathcal{P}(z,t) = 1-\tau - \frac{1-z}{1+t - t z}  = \frac{1}{(1+t)^2}\,\frac{z}{1-\tau z}
\end{eqnarray}
Expanding \eqref{gen-P:1} we recover \eqref{Pi:sol}. 

If $P_i(0) = \delta_{i,k}$, we get $g_0(z)=z^k$ and
\begin{equation}
\label{P-k:sol}
\mathcal{P}(z,t) = \left(1-\frac{1-z}{1+t - t z}\right)^k - \tau^k
\end{equation}

There are no infected with probability
\begin{equation}
\label{P0-k}
P_0 = g_0(\tau) = \tau^k\,, \quad \tau=\frac{t}{1+t}
\end{equation}
Generally the probabilities $P_i(t)$ are obtained by expanding \eqref{P-k:sol} in powers of $z$. Explicit formulas 
\begin{equation*}
\begin{split}
(1+t)^{k+1} P_1 &= \binom{k}{1}\,t^{k-1}\\
(1+t)^{k+2} P_2 &=  \binom{k}{1}\,t^{k}+ \binom{k}{2}\,t^{k-2}\\
(1+t)^{k+3} P_3 &= \binom{k}{1}\,t^{k+1} + 2\binom{k}{2}\,t^{k-1}+ \binom{k}{3}\,t^{k-3}
\end{split}
\end{equation*}
for $i=1,2,3$ help one to notice the pattern and suggest the general expression
\begin{equation}
\label{Pi-k}
P_i = \sum_{a=1}^i \binom{k}{a} \binom{i-1}{a-1}\,\frac{t^{k+i-2a}}{(1+t)^{k+i}}
\end{equation}
which can indeed be extracted by expanding \eqref{P-k:sol}.


\begin{thebibliography}{99}

\bibitem{McN}
     M.~McNeill, {\em Plagues and People}
     (Anchor Books, New York, 1989).

\bibitem{O}
     M.~Oldstone,
     {\em  Viruses, Plagues, and History}
     (Oxford University Press, Oxford, 1998).

\bibitem{Benedict}
     O. J. Benedictow, {\em The Black Death 1346--1353: The Complete History} (Boydell Press, Wiltshire, 2012). 

\bibitem{Ber}
     D. Bernoulli, M\'{e}m. Math. Phys. Acad. Roy. Sci., Paris, 1 (1760). 

\bibitem{Ber-history}
    K. Dietza and J. A. P. Heesterbeek, Math. Biosciences {\bf 180}, 1 (2002). 

\bibitem{MT}
     D.~P.~Maki and M.~Thompson, 
     {\em Mathematical Models and Applications, with emphasis on the
     social, life, and management sciences}
     (Englewood Cliffs, N.J., Prentice-Hall, 1973).

\bibitem{Lud}
     M.~A.~Ludwig,
     {\em The Giant Black Book of Computer Viruses}
     (American Eagle Publications Inc., Show Low, 1998). 

\bibitem{CFL}
     C. Castellano, S. Fortunato, and V. Loreto, Rev. Mod. Phys. {\bf} 81, 591 (2009).

\bibitem{Volovik} 
      P. L. Krapivsky, S Redner, and D. Volovik, J. Stat. Mech. P12003 (2011). 

\bibitem{McK}
     A.~G.~McKendrick, Proc.\ Edin.\ Math.\ Soc. {\bf 14}, 98 (1926).

\bibitem{KMcK}
     W. O. Kermack and A.~G.~McKendrick, Proc.\ Roy.\ Soc. A {\bf 115}, 700 (1927).

\bibitem{May}
     R.~Anderson and R.~May,
     {\em Infectious Diseases: Dynamics and Control}
     (Oxford University Press, Oxford, 1991).

\bibitem{Siam}
     H.~W.~Hethcote, 
     SIAM Rev. {\bf 42}, 599 (2000). 

\bibitem{Murray}
     J.~D.~Murray, {\em Mathematical Biology. I. An Introduction}
     (Springer-Verlag, New York, 2002).

\bibitem{Bailey50}
     N.~T.~J.~Bailey, Biometrika {\bf 37}, 193 (1950);
     {\it ibid} {\bf 40}, 177 (1953).

\bibitem{Bailey}
     N.~T.~J.~Bailey,
     {\em The Mathematical Theory of Infectious Diseases}
     (Oxford University Press, Oxford, 1987).

\bibitem{AB}
     H.~Andersson and T.~Britton,
     {\em Stochastic Epidemic Models and Their Statistical Analysis}
     (New York, Springer, 2000).

\bibitem{book}   
    P. L. Krapivsky, S. Redner and E. Ben-Naim,  {\it  A
    Kinetic View of Statistical Physics} (Cambridge, Cambridge University Press, 2010).

\bibitem{rr}     
     C.~J.~Ridler-Rowe,
     J. Appl.\ Prob. {\bf 4}, 19 (1967).

\bibitem{ML}
     A. Martin-L\"{o}f,  J. Appl. Probab. {\bf 35}, 671 (1998). 
         
\bibitem{bk}
     E.~Ben-Naim and P.~L.~Krapivsky,  
     Phys.\ Rev.\ E {\bf 69}, 050901(R) (2004).

\bibitem{KS}
     D.~A.~Kessler and N.~M.~Shnerb,  
     Phys.\ Rev.\ E {\bf 76}, 010901(R) (2007).

\bibitem{Gordillo}
     L. F. Gordillo, S. A. Marion,  A. Martin-L\"{o}f,  and P. E. Greenwood, Bull. Math. Biol. {\bf 70}, 589 (2008). 

\bibitem{Hofstad}
     R. van der Hofstad, A. J. E. M. Janssen, and J. S. H. Leeuwaarden, Adv. Appl. Probab. {\bf 42}, 706 (2010).
          
\bibitem{bk12}
     E.~Ben-Naim and P.~L.~Krapivsky,  Eur. Phys. J. B {\bf 85}, 145 (2012). 

\bibitem{Ginestra}
     F. Radicchi and G. Bianconi,  arXiv:2007.15034. 

\bibitem{feller} 
    W. Feller,  {\em An Introduction to Probability Theory and Its Applications}, 
    Vol. I, 3rd edn. (John Wiley, New York, 1968).

\bibitem{teh}
     T.~E.~Harris,
     {\em The Theory of Branching Processes} (Dover, New York, 1989).

\bibitem{athreya04} 
    K. B. Athreya and P. E. Ney, {\em Branching Processes} 
    (Dover Publications, Inc., Mineola, New York, 2004).

\bibitem{branch} 
    M. Kimmel and D. Axelrod, {\it Branching Processes in Biology} 
    (Springer, New York, 2002).
    
\bibitem{vatutin} 
    P. Haccou, P. Jagers, and V. A. Vatutin, 
    {\it Branching processes: variation, growth, and extinction of populations} 
    (Cambridge University Press,  New York, 2005).

\bibitem{Knuth} 
     R.~L.~Graham, D.~E.~Knuth, and O.~Patashnik, 
     {\em Concrete Mathematics: A Foundation for Computer Science}
     (Reading, Mass.: Addison-Wesley, 1989).
    
\bibitem{caveat}
    The same leading behavior \eqref{time-av-log}  is obtained if one would use the upper bound $C\sqrt[3]{N}$ with arbitrary $C$; due
    to the logarithmic divergence of the integral, only the scaling of the upper bound with $N$ is required for extracting the 
     leading behavior. 

\bibitem{Satya} 
     S. N. Majumdar and R. M. Ziff, Phys.\ Rev.\ Lett. {\bf 101}, 050601 (2008).
    
    
\end{thebibliography}
\end{document}